\documentclass[prl,twocolumn,showpacs,amsmath,superscriptaddress,preprintnumbers,amssymb]{revtex4-2}
\usepackage{graphicx}
\usepackage{xcolor}
\usepackage{dcolumn}
\usepackage{bm}
\usepackage{wasysym}

\newcommand{\beq}{\begin{equation}}
\newcommand{\eeq}{\end{equation}}
\newcommand{\beqn}{\begin{eqnarray}}
\newcommand{\eeqn}{\end{eqnarray}}

\newcommand{\ra}{\rightarrow}

\newcommand{\cC}{ {\cal C} }
\newcommand{\cT}{ {\cal T} }

\newcommand{\cL}{ {\cal L} }

\newcommand{\vect}[1]{{\bm{#1}}}

\newcommand{\ii}{\mathrm{i}}

\begin{document}

\title{The Conjugate Composite Fermi Liquid }

\author{Nayan Myerson-Jain}

\affiliation{Department of Physics, University of California, Santa Barbara, CA 93106}

\author{Chao-Ming Jian} \affiliation{Department of Physics, Cornell University, Ithaca,
New York 14853, USA}

\author{Cenke Xu}

\affiliation{Department of Physics, University of California, Santa Barbara, CA 93106}

\begin{abstract}

Recent experimental observations of the fractional quantum anomalous Hall effect in spin/valley polarized moir\'{e} systems call for a more expansive theoretical exploration of strongly correlated physics in partially filled topological bands. In this work we study a state that we refer to as the conjugate-composite Fermi liquid (cCFL), which arises when a pair of Chern bands with opposite Chern numbers are both half-filled. We demonstrate that the cCFL is the parent state of various interesting phenomena. As an example, we demonstrate that with the existence of an inplane spin order, the cCFL could be driven into a quantum bad metal phase, in the sense that it is a metallic phase whose zero temperature longitudinal resistivity is finite, but far greater than the Mott-Ioffe-Regal limit, i.e. $\rho^e_{xx} \gg h/e^2$. The bad metal phase is also accompanied with a new Wiedemann-Franz law, meaning the thermal conductivity is proportional to the electrical resistivity rather than conductivity. Other proximate phases of the cCFL such as superconductivity and a chiral spin liquid phase can occur when the composite fermions (CF) form the inter-valley CF-exciton condensate.

\end{abstract}

\date{\today}

\maketitle

{\bf --- Introduction}

The fractional Chern insulator~\cite{youngFCI}, especially the fractional quantum anomalous Hall (FQAH) effects predicted~\cite{linFQAH,fuFQAH1} and observed ~\cite{xuFQAH1,xuFQAH2,makFCI,juFQAH} at zero external magnetic field in diverse moir\'{e} systems represent a milestone of the study of the interplay between nontrivial band topology and strongly correlated physics, and they have motivated a series of recent theoretical studies of the FQAH-related phases and nearby phase transitions~\cite{cano,ashvinCFL,fuCFL,senthilFQAH1,bernevig,zhangFQAH,songsenthil,xufuFQAH}. The FQAH effect necessarily requires time-reversal symmetry breaking and valley polarization, and the most relevant physics occurs at one partially filled Chern band with a nonzero Chern number. Nontrivial correlated physics can also emerge in a pair of Chern bands with opposite Chern numbers. When spin-up and spin-down electrons each occupy a Chern band with opposite Chern numbers, the system forms a quantum spin Hall insulator~\cite{QSH1,QSH2}. A quantum spin Hall insulator is not only an archetypal example of a topological insulator, it is also a platform of rich exotic physics, such as charged-skyrmion condensation induced superconductivity~\cite{grover2008topological,ashvinskyrmion}. Previous theoretical studies have also explored incompressible topological states when both spin-up and down electrons partially fill a pair of conjugate Chern bands, and it was shown that interesting topological defects with nonabelian statistics can be engineered based on these conjugate topological states~\cite{Barkeshli_2013,cheng2012}.

However, novel {\it compressible} states that emerge from partially filled conjugate Chern bands have been rarely explored. The most well-known nontrivial compressible state that arises from the quantum Hall physics is the composite Fermi liquid (CFL) that occurs at a half-filled Landau level, which may be described by either the HLR theory~\cite{HLR}, or the Son's QED$_3$ theory~\cite{Son2015}. Although the HLR state and the Son's state were both proposed for the half-filled Landau level, recent studies have indicated that physics of a partially-filled Chern band (at least for Chern bands with Chern number $\pm 1$ in the context of TMD moir\'{e} system)~\cite{macdonaldLL,Paul_2023} can be understood as a partially filled Landau level, i.e. the Chern band with Chern number $\pm 1$ at filling factor $\nu$ can be viewed as a Landau level filled with electron density $n_e = \pm \nu n_{\phi}$, where $n_\phi$ is the density of magnetic flux. 

In this work, we explore the physics of a pair of CFLs that are conjugate to each other in the sense that they arise from half-filled Chern bands with opposite Chern numbers, as well as its proximate phases. We will demonstrate that the cCFL state serves as the parent state of several interesting proximate phases, including an exotic ``quantum bad metal" phase, a superconductor, and a chiral spin liquid phase with four fold topological degeneracy.

{\bf --- The conjugate composite Fermi liquid}

Let us start with the homobilayer transition metal dichalcogenide (TMD) moir\'{e} system with spin-valley locking as an example. According to Ref.~\cite{wu2019topological},  within a certain twisting angle the moir\'{e} minibands are analogous to those of a quantum spin Hall insulator. The underlying system has a time-reversal $\mathcal{T}$ symmetry.
The most relevant bands we consider is the Chern band with Chern number $C_1 = +1$ from one spin/valley, and another band with Chern number $C_2 = -1$ from the opposite spin/valley. We consider the situation when the electrons half-fill ($\nu = 1/2$) both bands mentioned above. At half-filling, the system at each valley may form a composite Fermi liquid (CFL), which was seen in recent numerics that models physics at a single valley of the TMD moir\'{e} system~\cite{ashvinCFL,fuCFL}.
The CFL can be described by either the HLR state~\cite{HLR}, or the Son's QED$_3$ state~\cite{Son2015,wanghall2}. 
We will use the QED$_3$ state as an example. When both spin/valleys form Son's QED$_3$ states and they are connected to each other through $\mathcal{T}$, the (simplified) Lagrangians for the two valleys read \beqn \cL_{1} = \cL_D(\psi_1, a_1) - \mu \psi^\dagger_1 \psi_1 - \frac{1}{2} \frac{\ii}{2\pi} a_1 d A_1 - \frac{1}{2} \frac{\ii}{4\pi} A_1 d A_1; \cr\cr \cL_{2} = \cL_D(\psi_2, a_2) + \mu \psi^\dagger_2 \psi_2 - \frac{1}{2} \frac{\ii}{2\pi} a_2 dA_2 + \frac{1}{2} \frac{\ii}{4\pi} A_2 d A_2, \label{LSon} \eeqn where $\cL_D$ is the Lagrangian for the Dirac composite fermion (CF) $\psi_{1,2}$ which is minimally coupled to gauge fields $a_{1,2}$, and $adA$ is an abbreviated notation of $a \wedge dA$. The valley mixing is expected to be very weak with small twisting angle and the electric charges in both spin/valleys are conserved separately, hence we have introduced an external gauge field $A_{1,2}$ for each spin/valley. It is worth noting that spin transport and spin accumulation has been observed in TMD moire systems~\cite{spinhallTMD}, meaning that the electrons in each spin/valley are indeed well conserved separately in these systems. 

The dynamical gauge field $a_i$ of the QED$_3$ state is the dual of the electron current in each spin/valley: $J^e_{i,\mu} \sim \epsilon_{\mu\nu\rho} \partial_\nu a_{i,\rho}$. Hence the Dirac CFs $\psi_{i}$ that couple to $a_{i,\mu}$ are vortices of electric charges. Since the vorticity changes sign under $\mathcal{T}$, the Dirac CF should undergo a particle-hole transformation under $ \mathcal{T}: \psi_{1,2} \rightarrow \psi^\dagger_{2,1}$. The chemical potential of the Dirac CF is nonzero generally and the two spin/valley flavors should have opposite signs of the chemical potential, if we assume the entire system is $\mathcal{T}$-invariant. Here, the CS terms in Eq.~\ref{LSon} are not properly quantized, but it can be easily remedied. The remedied form with correct quantization will become important when we discuss gapped proximate phases near the cCFL.


{\bf --- The quantum bad metal phase}

For two-dimensional systems, the electrical resistivity (and conductivity) is a dimensionless quantity, hence it could be universal in the unit of $h/e^2$. For example, it was pointed out that the critical resistivity at a $(2+1)d$ QCP will take a universal value at the order of $h/e^2$~\cite{fisher1990coulomb}. The electrical conductivity in the AC limit captures the universal infrared physics of the QCP, and is often studied together with the central charge of the corresponding conformal field theory~\cite{Giombi_2016,Diab_2016,PhysRevB.88.155109}. In noninteracting $2d$ electron systems, depending on the symmetry class (such as sympletic) there could be a metal-insulator transition due to localization, depending on the strength of disorder, and the critical resistivity would also be at the order of $h/e^2$~\cite{andersonreview}. In fact, for the rudimentary methods of charge transport to be applicable, one would need $k_F l_{\mathrm{m}} \gg 1$, where $k_F$ and $l_m$ are the Fermi wave vector and the mean free path, which sets an upper bound for the resistivity of a metal, the so-called Mott-Ioffe-Regal (MIR) limit. When $k_F l_{\mathrm{m}} \sim 1$, the resistivity of the $2d$ system is at the order of $h/e^2$. Hence if a $2d$ metal has a resistivity far greater than $h/e^2$, it is considered as a bad metal~\cite{bad_metal}.

A bad metal is a notion that was often discussed in the context of the cuprates materials, where the resistivity of the strange metal phase increases linear with temperature $T$, and it exceeds the MIR limit at certain finite temperature. In this work we will demonstrate that the cCFL is a parent state on which we can construct a ``quantum bad metal phase", which means that it is a metallic phase with finite longitudinal resistivity at {\it zero temperature}, but the resistivity at zero temperature far exceeds the MIR limit. Upon raising temperature, the resistivity of the system would actually decrease. It is worth noting that, behaviors similar to the quantum bad metal described above have been observed in a heterobilayer TMD moir\'{e} system without band topology~\cite{tmdmit1}: it was observed that at a bandwidth-tuned metal-insulator transition, the critical resistivity at zero temperature is far greater than $h/e^2$, and it decreases with increasing temperature. Theories for this anomalous metal-insulator transition were proposed~\cite{fractionalMIT,senthilmit3} based on different pictures from the current work.

To evaluate the transport properties of the cCFL state, let us first just investigate one of the two valleys. The electrical conductivity tensor for the QED$_3$ CFL state reads~\footnote{This is the same composition rule as derived in Ref.~\cite{wanghall2}, but it is a different composition rule as the HLR state, where the electrical resistivity is the sum of the CF-resistivity and that from the background gauge field $A_1$.} \beqn \sigma^e_{1} = \frac{1}{4} (\sigma^{\mathrm{cf}}_{1})^{-1} + \Sigma_{A,1}, \label{condson} \eeqn where $\sigma^{\mathrm{cf}}_{1}$ is the conductivity tensor for the Dirac CF in spin/valley-1, and $\Sigma_{A,1}$ originates from the level-$-1/2$ CS term of the background electromagnetic field $A_1$: \beqn
\sigma^{\mathrm{cf}}_{1} =
\begin{pmatrix}
\sigma^{\mathrm{cf}} & 0 \\
0 & \sigma^{\mathrm{cf}}
\end{pmatrix}; \ \ \
\Sigma_{A,1} =
\begin{pmatrix}
0 & + 1 /2 \\
- 1 /2 & 0
\end{pmatrix}; \eeqn which leads to the electrical conductivity tensor
\beqn \sigma^e_{1} =
\begin{pmatrix}
1/(4\sigma^{\mathrm{cf}}) & + 1 /2 \\ \\
- 1 /2 & 1/(4\sigma^{\mathrm{cf}}) \label{condson2}
\end{pmatrix}
\eeqn
Since the Dirac CF is physically the vortex of charge, intuitively one would expect that a good conductor of CF means a bad metal of charge. Indeed, it was proposed that the product of the vortex conductivity and the electrical conductivity is a constant~\cite{fisher1990coulomb}. Hence let us assume that the longitudinal conductivity of the CF $\sigma^{\mathrm{cf}}$ is large, which is a reasonable assumption when there is a finite Fermi surface of $\psi$ and long mean free path due to weak disorder. With a large longitudinal conductivity of the CF $\sigma^{\mathrm{cf}}$, the longitudinal electrical conductivity is indeed small, as $\sigma^e_{xx} \sim 1/ (4 \sigma^{\mathrm{cf}})$. This seems consistent with the naive intuition, as a good conductor of the vortices (i.e. the CF) should be a bad conductor of electric charges. However, the relation $\sigma^e_{xx} \sim 1/ (4 \sigma^{\mathrm{cf}})$ does not immediately lead to a large longitudinal electrical resistivity, which is what is often measured experimentally. Due to the existence of transverse component of the conductivity tensor, inverting the conductivity tensor leads to the longitudinal electrical resistivity \beqn \rho^e_{xx} = \frac{4 \sigma^{\mathrm{cf}}}{1 + 4 (\sigma^{\mathrm{cf}})^2 }. \eeqn In the limit of large $\sigma^{\mathrm{cf}}$, $\rho^e_{xx}$ approaches zero.

The same computation can be carried out for the HLR state for valley-1. However, there is a subtlety that the transverse transport of the CFs of the HLR state cannot be ignored, and its value is mandated if one assumes an exact particle-hole symmetry of the half-filled Landau level. This was pointed out by Ref.~\cite{kivelsonhlr} and explicitly showed in Ref.~\cite{wanghlr}. 
Once we correctly take into account the transverse transport of the CFs, the calculation based on the HLR state will yield consistent results as the QED$_3$ state for the CFL.

The reason for the longitudinal resistivity to remain small is due to the transverse transport, and hence it appears that as long as we consider the two valleys together, the transverse transport would cancel out due to the time-reversal conjugation between the two valleys. However, simply combining the two valleys would not help to get a large longitudinal electrical resistivity, as there is still a transverse spin current which contributes to the overall resistivity and prohibits the desired bad metal behavior. Upon combining the two valleys, the gauge-fields for each valley $A_1,A_2$ can be recast into the electromagnetic gauge field $A^e = \frac{1}{2}(A_1+A_2)$ and the ``spin gauge field" $A^s = A_1 - A_2$. For the combined cCFL state the total conductivity tensor for the combined system is \beqn \sigma^e_{\text{total}} &=&
\frac{1}{2} (\tilde{\sigma}^{\mathrm{cf}})^{-1} +
\Sigma^{e,s}_{A}, \cr\cr \frac{1}{2}
(\tilde{\sigma}^{\mathrm{cf}})^{-1} &=&
\begin{pmatrix}
1/(2 \sigma^\text{cf}) &0 &0 & 0 \\
0 & 1/(2 \sigma^\text{cf}) & 0 & 0\\
0 & 0 &1/(8 \sigma^\text{cf}) &  0\\
0 & 0 & 0 & 1/(8 \sigma^\text{cf})
\end{pmatrix}; \cr\cr\cr
\Sigma^{e,s}_{A} &=&
\begin{pmatrix}
0 &0 &0 & + 1/2 \\
0 & 0 & -1/2 & 0\\
0 & +1/2 & 0 &  0\\
-1/2 & 0 & 0 & 0
\end{pmatrix} \label{totalcond}
\eeqn in the basis $(A^e_x, A^e_y, A^s_x, A^s_y)$. The total resistivity tensor for the system is now easily seen to be \beqn && \rho^e_{\text{total}} = \cr\cr && \frac{1}{1+4(\sigma^\text{cf})^2}
\begin{pmatrix}
2 \sigma^\text{cf}&0&0&-8(\sigma^\text{cf})^2\\
0&2\sigma^\text{cf}&8(\sigma^\text{cf})^2&0\\
0&-8(\sigma^\text{cf})^2&8  \sigma^\text{cf}&0\\
8(\sigma^\text{cf})^2&0&0&8  \sigma^\text{cf}
\end{pmatrix} .
\eeqn Due to the spin hall effect, the longitudinal electrical resistivity is $\rho^e_{xx,\text{total}} = 2 \sigma^\text{cf}/(1+4(\sigma^\text{cf})^2)$, which still remains small as $\sigma^\text{cf} \rightarrow \infty$.

The physical picture of this result is that, when there is an electrical current along the $\hat{x}$ direction, it will predominantly induce a spin-voltage drop along the $\hat{y}$ direction rather than a electrical voltage drop along the $\hat{x}$ direction, due to the mutual charge-spin Hall effect. To remedy this, we consider the scenario that the system develops an inplane spin order. Here we assume that the spins of electrons in each valley is polarized along the $\hat{z}$ direction, and an inplane spin order is a ``superfluid" phase of spin $S^z$. The effect of the inplane spin order can be taken into account by introducing the spin conductivity $\sigma^s$ into $\Sigma^{e,s}_{A}$, and taking the limit $\sigma^s \rightarrow \infty$: \beqn \Sigma^{e,s}_{A} =
\begin{pmatrix}
0 &0 &0 & + 1/2 \\
0 & 0 & -1/2 & 0\\
0 & +1/2 & \sigma^s &  0\\
-1/2 & 0 & 0 & \sigma^s \label{Sigma2}
\end{pmatrix}. \eeqn In the limit of $\sigma^s \ra \infty$, repeating the calculation above leads to longitudinal electrical resistivity \beqn \lim_{\sigma^s \ra \infty} \rho^e_{xx} = 2\sigma^{\mathrm{cf}}. \eeqn Now assuming a good metal of the CFs would indeed imply a bad metal of the electric charges. Here we need the momentum of the inplane spin-order to be commensurate with the two valleys, in order to Higgs the spin/valley gauge field $A^s$. A background inplane spin order that connects the two valleys will involve monopole operators of the dynamical gauge fields in the Son's QED$_3$ theory, but, it is commonly expected that when there is a finite Fermi surface of the matter fields coupled to the gauge field, the monopole events would be irrelevant~\cite{leemonopole}.

The relaxation of the CF current can result from scattering with the disorder, and the dynamical gauge fields. At very low temperature, the gauge fields are suppressed, hence the CF conductivity $\sigma^{\mathrm{cf}}$ can be large at low temperature due to the existence of a finite CF Fermi surface, when the disorder is weak enough. Then the electrical resistivity can be very large but finite (in the unit of $h/e^2$) at low temperature. While increasing temperature $T$, the CF conductivity would naturally decrease due to enhanced scattering from the thermally activated gauge fields, hence {\it the electrical resistivity would also decrease with increasing temperature.} The $T-$dependence of the CF transport can be evaluated in the same way as the standard calculation for the spinon Fermi surface coupled with a dynamical gauge field~\cite{leenagaosa}. More specifically, we expect that $\rho^e_{xx}(T)$ decreases with temperature as \beqn \rho^{e}_{xx}(T) \sim \rho^{e}_{xx}(0) - c T^{4/3}, \eeqn at low temperature with a constant positive $c$.

Other peculiar properties of bad metal phase can be deduced. For example, the thermal transport of the system is still dominated by the composite fermion, and there should be a ``Wiedemann-Franz" law relating the thermal conductivity and the composite fermion conductivity: $\kappa_{xx} = L_{\mathrm{cf}} T \sigma^{\mathrm{cf}}$, where $L_{\mathrm{cf}}$ is proportional to $(e^2/\hbar) L_0$, and $L_0$ is the Lorenz ratio of free electrons (we have taken the CF conductivity $\sigma^{\mathrm{cf}}$ as a pure number). Then since $\rho^e_{xx} = 2 \sigma^{\mathrm{cf}}$ in units of $h/e^2$, the Wiedemann-Franz law for this quantum bad metal reads \beqn \kappa_{xx} = \frac{e^2}{2h} L_{\mathrm{cf}} T \rho^e_{xx}. \eeqn This is a strong violation of the standard Wiedemann-Franz law of the Fermi liquid, and it is a different violation from the result in the ordinary CFL~\cite{wanghall1}. In the supplementary material we will discuss transport of the bad metal under a canted magnetic order.

Instead of the scenario with an in-plane spin order, another interesting scenario concerns the cCFL under an in-plane magnetic field. Due to the momentum mismatch of the two valleys, a uniform in-plane magnetic field does not relax the spin current and the $S_z$ quantum number in the bulk. However, the $S_z$ is no longer conserved near the boundary where the momentum is not conserved, and the two valleys are allowed to hybridize (under the in-plane field). Hence, in a transport measurement of the cCFL under an in-plane field, one should treat the system's boundary as a source/drain of the spin current. Now, consider the transport measurement in a setting where the electric current runs along the $x$ direction, and the system has a finite size along the $y$ direction. Due to the boundaries, the system is expected to have a vanishing spin voltage drop along the $\hat{y}$ direction (while the spin current along the $\hat{y}$ direction can be finite). Following Eq. \eqref{totalcond}, the longitudinal resistivity measured in this setting, defined by the ratio between the electric voltage drop along the $x$ direction and the electric current, is given by $\rho^{e'}_{xx} = 2 \sigma^{\rm cf}$, which exhibits a quantum bad metal behavior.

{\bf --- Superconductor}

Hereafter we explore the gapped proximate descendant phases that can be naturally constructed from the cCFL. The cCFL is a gapless parent phase, and gapped descendant phases near the cCFL can be constructed by (for example) forming the exciton condensate of the CFs. Intuitively, since the CFs of the Son's QED$_3$ state are vortices of charge, if the CFs form a trivial gapped insulator without band topology through the CF-exciton condensate, the system should naturally develop superconductivity according to the standard particle-vortex duality in $(2+1)d$~\cite{peskindual,halperindual,leedual}. If the CF-exciton leads to a band insulator of the CF with nontrivial band topology instead, the system may develop topological order, as we shall see in the next section. 

Since these descendant states are gapped, we need to be careful with the proper quantization of the CS terms in Eq.~\ref{LSon}~\cite{SEIBERG2016}. The completed theory of Eq.~\ref{LSon} reads: \beqn \cL_{1} = \cL_{+\frac{1}{2}}(\psi_1, a_1, \mu) 
+ \frac{\ii}{2\pi} a_1 d b_1 + \frac{2\ii}{4\pi} b_1 d b_1 + \frac{\ii}{2\pi} b_1 d A_1; \cr\cr \cL_{2} = \cL_{-\frac{1}{2}}(\psi_2, a_2, - \mu) 
- \frac{\ii}{2\pi} a_2 d b_2 - \frac{2\ii}{4\pi} b_2 d b_2 + \frac{\ii}{2\pi} b_2 d A_2. \cr \label{LSon2} \eeqn The Maxwell terms of the dynamical gauge fields $a_i$, $b_i$ are not written explicitly. We note that gauge fields $a_i$ and $b_i$ transform in the same way under $\mathcal{T}$, but oppositely from the external gauge fields $A_i$. Here $\cL_{\pm 1/2}$ is a single Dirac fermion coupled with a gauge field with a level-$\pm 1/2$ CS term, but the proper form of $\cL_{\pm 1/2}$ would involve the $\eta-$invariant of the gauge field. $\cL_{\pm 1/2}$ can be viewed as fermion $\psi_i$ coupled with gauge field $a_i$, when the filled band of $\psi_i$ is tuned to the transition between a Chern insulator and a trivial insulator, i.e. there is another gapped heavy Dirac fermion in the band structure that would formally lead to the level-$\pm 1/2$ CS term of $a_i$. 
If $\psi_1$ and $\psi_2$ form an inter-valley CF-exciton condensate, i.e. $\langle \psi^\dagger_1 \psi_2 \rangle \neq 0$, this would drive a Higgs transition and identify $a_1 = a_2 = a$. Here the CF-exciton condensate would lead to a fully gapped spectrum of the CFs. In this section we need this gapped band structure to have trivial topology in total, hence there is no CS term of $a$ generated after integrating out all the CFs.

Integrating out gauge field $a$ leads to $b_1 = b_2 = b$, then the Lagrangian $\cL = \cL_1 + \cL_2 $ reduces to the effective form \beqn \cL_{\mathrm{eff}} = \frac{\ii}{2\pi} b d(A_1 + A_2) + \cdots = \frac{2 \ii}{2\pi} b d A^e + \cdots. \eeqn This coupling means that, the flux of $b$ carries two units of charges of $A^e$. And since the charge is conserved, the monopole effect that creates/annihilates the flux of $b$ is suppressed, and $b$ would be in a photon phase described by a Maxwell term in the ellipsis. The photon phase of $b$ will be dual to the condensate of the flux of $b$, which means that the exciton condensate of the CFs is actually a superconducting phase of charges with charge$-2e$ Cooper pair.

To realize the superconductivity through the CF-exciton condensation described above, one needs to turn on a repulsive interaction between the CFs of the two valleys: $H^{\mathrm{int}} = U \psi^\dagger_1 \psi_1 \psi_2^\dagger \psi_2$. 
One natural question is whether the cCFL is unstable against an infinitesimal perturbation of positive $U$. In the following we will argue it is likely that the repulsive interaction needs to be above certain critical value $U_c$ for the system to develop superconductivity. The reason is that, $\psi_{1,2}$ are coupled with the gauge field $a_{1,2}$ respectively, and the flux of $a_{1,2}$ are both dual to the electric charge density, hence there is naturally a repulsive interaction between the fluxes of $a_1$ and $a_2$: $g (\vect{\nabla} \times \vect{a}_1)(\vect{\nabla} \times \vect{a}_2)$, with $g > 0$. One can recombine the Maxwell terms of $\vect{a}_1$ and $\vect{a}_2$ into $(\vect{\nabla} \times \vect{a}_+)^2/e_+^2 + (\vect{\nabla} \times \vect{a}_-)/e_-^2 $. The CFs $\psi_1$ and $\psi_2$ carry the same charge under $a_+$ but {\it opposite} charge under $a_-$, hence $a_+$ and $a_-$ would respectively generate a repulsive and attractive interaction between $\psi_1$ and $\psi_2$. A positive $g$ implies that $e_+ < e_-$, meaning that the inter-valley interaction between the CFs generated by the gauge fields are likely attractive, and hence the interaction $U$-term needs to be greater than a critical value $U_c$ to induce an inter-valley CF-exciton condensate. A more careful renormalization group analysis of $U$ with the presence of the interaction between the gauge fluxes can be performed in the same manner as Ref.~\cite{zoudebanjan,mandal}.

{\bf --- Chiral spin liquid}

A chiral spin liquid state (with a background integer quantum hall state) is also a natural correlated state for a pair of half-filled conjugate Chern bands. Let us imagine that the spin/valley-2 is first fully filled with electrons (which would form an integer quantum Hall state), then half-filled with holes. The electrons in spin/valley-1 and holes in spin/valley-2 will effectively see the same magnetic flux density: $n_{\phi} = 2 n_{e,1} = 2 n_{h,2}$. Then the inter-valley exciton, a bound state between electron and hole, will be at $1/4$ filling compared with the effective magnetic flux density $n_\phi^\ast$ seen by the bound state: $n_{ex} = n^\ast_{\phi}/4$. The inter-valley exciton carries spin $S^z = 1$ as well. Hence it is natural to expect that the spins can form a chiral spin liquid state that is analogous to the bosonic fractional quantum Hall state at $\nu = 1/4$, which is similar to the renowned chiral spin liquid first constructed in Ref.~\cite{laughlinsl}, except that now the chiral spin liquid has four-fold ground state degeneracy. In terms of a CS theory, this state should be described by the following Lagrangian: \beqn \cL_{\mathrm{CSL}} = \frac{4\ii}{ 4\pi } b db + \frac{\ii}{2\pi} b d A^s - \frac{\ii}{4\pi} cdc + \frac{\ii}{2\pi} cdA_2. \label{CSL} \eeqn The last two terms represent the background integer quantum Hall state arising from the fully filled valley-2 by electrons. 

To show that the chiral spin liquid state described above is also a proximate state of the cCFL, we can still form a CF-exciton condensate between the two valleys described by Eq.~\ref{LSon2}, but now we need the fully gapped bands of $\psi_i$ due to the CF-exciton condensate to have a total Chern number $+1$, through for example a $p + \ii p$-wave exciton condensate. The CF-exciton condensate not only identifies $a_1 = a_2 = a$, after integrating out the fully gapped CFs, a CS term will be generated for gauge field $a$: $- \frac{\ii }{4\pi} a d a$. Since $a$ is coupled with both $b_1$ and $b_2$, integrating out $a$ will lead to the following Lagrangian for $b_i$: \beqn \cL &=& \frac{3\ii}{4\pi}b_1 d b_1 - \frac{\ii}{4\pi} b_2 d b_2 - \frac{\ii}{2\pi} b_1 d b_2 \cr\cr &+& \frac{\ii}{2\pi} b_1 d A_1 + \frac{\ii}{2\pi} b_2 d A_2. \label{CSL2} \eeqn Now after a similar transformation (relabelling the gauge fields) $b_1 \ra b $, $b_2 \ra c - b$, Eq.~\ref{CSL2} reduces to Eq.~\ref{CSL}. One can also check that, if the CF-exciton induces a gapped CF band with total Chern number $-1$, the system would become the ``conjugate" chiral spin liquid state of Eq.~\ref{CSL}, i.e. the electrons in valley-1 form an integer quantum Hall state first, then the bound state of electron from valley-2 and hole from valley-1 would form a $\nu = -1/4$ FQAH state, which is the time-reversal conjugate of the chiral spin liquid in Eq.~\ref{CSL}. In the SM we will show that, exploiting the PH symmetry of the Son's QED$_3$ state, the chiral spin liquid has another description.

In fact, similar chiral spin liquids can be formed when one valley has electron filling factor $1/k$ with integer $k$, the other has electron filling factor $1 - 1/k$. The argument above would suggest that the inter-valley excitons have filling factor $1 /(2k)$ compared with the effective magnetic flux density they see. Hence, incompressible chiral spin liquids with bosonic filling factor $\nu = 1/(2k)$ can be formed.

{\bf --- Summary}

In this work we discussed a cCFL state that naturally emerges when a pair of Chern bands with opposite Chern numbers are both half-filled. We demonstrated that the cCFL state serves as the parent state of several desirable phases, including a quantum bad metal, a superconductor, and a chiral spin liquid. The composite fermion in the cCFL can be viewed as a type of charge vortex, and we would like to acknowledge that the rich physics of various types of ``vortex liquids" were discussed in previous works, motivated from different contexts ~\cite{wanghall1,wanghall2,mulligan1,mulligan2,vortexliquidfisher,philip,myersonjain2022vortex}. The cCFL state discussed in the current work can also be viewed as a particular type of vortex liquid. Given the rapid experimental progress on the FQAH related phenomena, the cCFL state discussed in this work and its descendant phases may become relevant to near-future experiments on moir\'{e} heterostructures. 

The cCFL considered in this work is a compressible state constructed from partially filled multiple Chern bands. We expect that abundant interesting phenomena can arise from {\it compressible} states with fractional filling factors on more than one Chern bands, especially on Chern bands with higher Chern numbers. We note that some explorations along this direction were conducted in Ref.~\cite{wanghall3}. We leave the general exploration of nontrivial compressible states arising from Chern bands to future study. 

C.X. is supported by the Simons Foundation through the Simons Investigator program. After we finished our manuscript, we became aware of a previous preprint~\cite{zhang2018composite}, where the authors also discussed CFL from half-filling a pair of conjugate Chern bands, with focus on a different set of descendant states.

\bibliography{cCFL01}

\appendix

\section{Bad metal in a canted spin order}

Let us now consider a canted spin order, with a small ferromagnetic moment along the $S^z$ direction. This will amount to causing imbalance of average charge density on the two valleys, and lead to a nonzero average internal magnetic field $\bar{b}\hat{z} = \vect{\nabla} \times \vect{\bar{a}}_1 = - \vect{\nabla} \times \vect{\bar{a}_2}$. In this case the CFs in each valley will gain a small
Hall resistivity. Note that although this turns on an internal magnetic field opposite in sign between the two valleys, since the group velocity of the CFs in valley$-2$ is opposite of those in valley-$1$ in the Son's QED$_3$ state, the CFs in each valley effectively see the same magnetic field and have the same Hall resistivity $\rho_{xy}^\text{cf}$.  In the total electron conductivity tensor $\sigma^e_\text{total} = \frac{1}{2}(\tilde{\sigma}^\text{cf})^{-1} + \Sigma_A^{e,s}$, the CF contribution is now modified to
\beqn
\frac{1}{2}
(\tilde{\sigma}^{\mathrm{cf}})^{-1} &=&
\begin{pmatrix}
1/(2 \sigma^\text{cf}) &\rho_{xy}^\text{cf}/2 &0 & 0 \\
-\rho_{xy}^\text{cf}/2 & 1/(2 \sigma^\text{cf}) & 0 & 0\\
0 & 0 &1/(8 \sigma^\text{cf}) &  \rho_{xy}^\text{cf}/8\\
0 & 0 & -\rho_{xy}^\text{cf}/8 & 1/(8 \sigma^\text{cf})
\end{pmatrix}
\eeqn
with $\Sigma^{e,s}_A$ in Eq.~\ref{Sigma2}, which contains the large contribution to the spin conductivity from ordering. The longitudinal and transverse components of the resistivity of the electrons in the limit of $\sigma^s \rightarrow \infty$ in Eq.~\ref{Sigma2} now become
\beqn
\rho^e_{xx} = \frac{2 \sigma^\text{cf}}{1 + (\sigma^\text{cf} \rho_{xy}^\text{cf})^2 }, \text{ }
\rho^e_{xy} = -\frac{2(\sigma^\text{cf})^2 \rho^\text{cf}_{xy} }{1 + (\sigma^\text{cf} \rho_{xy}^\text{cf})^2} \label{CantedCond}.
\eeqn
The value of the longitudinal resistivity has been reduced from the original case. Under imbalance of charge density between the two valleys, which breaks the time-reversal symmetry of the system, the electrons gain an anomalous Hall response as well. 

Alternatively, if we would like to dope charge by turning on a magnetic field for $\vect{a}_+$ (magnetic fields in the same direction for each valley) the CFs in valley$-1$ see an opposite Hall resistivity from those in valley-$2$. The CF contribution to the total electron conductivity tensor under this kind of doping is 
\beqn
\frac{1}{2}
(\tilde{\sigma}^{\mathrm{cf}})^{-1} &=&
\begin{pmatrix}
1/(2 \sigma^\text{cf}) &0 &0 & \rho_{xy}^\text{cf}/4 \\
0 & 1/(2 \sigma^\text{cf}) & -\rho_{xy}^\text{cf}/4 & 0\\
0 & \rho_{xy}^\text{cf}/4 &1/(8 \sigma^\text{cf}) &  0\\
-\rho_{xy}^\text{cf}/4 & 0 & 0 & 1/(8 \sigma^\text{cf}).
\end{pmatrix}.
\eeqn
The longitudinal and transverse components of the electron resistivity remain unchanged, i.e. $\rho^e_{xx} = 2 \sigma^\text{cf}$ and $\rho^e_{xy} = 0$. The quantum bad metal phase remains robust under small charge doping. 

Diffusive behavior of CFs near the Fermi surface, which lends to the dominant contribution to electronic transport quantities such as $\sigma^\text{cf}$ and $\rho_{xy}^\text{cf}$, is well-modeled by semiclassical dynamics. The standard semiclassical expression for the DC conductivity tensor for CFs in valley-1 in the presence of a magnetic field $\bar{b}$ is
\beqn
\sigma^{\text{cf}}_{1,ij} \sim  \tau \int d^2q \text{ } v_i(\vect{q}) \langle v_j(\vect{q}) \rangle \delta(\epsilon(\vect{q}) - \mu ) \label{semiclassicalcond}.
\eeqn
We have taken the charge of the CFs to be one. The Dirac CFs of Son's parent QED$_3$ state exhibit a linear dispersion $\epsilon(\vect{q}) = v_F |\vect{q}|$, with group velocity $\vect{v}(\vect{q}) = \nabla_\vect{q} \epsilon(\vect{q})$. The quantity $\langle \vect{v}(\vect{q}) \rangle$ is the group velocity time-averaged over its collision history,
\beqn
\langle \vect{v}(\vect{q}) \rangle = \int_{-\infty}^0 \frac{dt}{\tau} \text{ }  e^{t/\tau} \vect{v}(\vect{p}(t)) \label{avevel}.
\eeqn
The momentum $\vect{v}(\vect{p}(t))$ is a solution to the semiclassical equation of motion in the presence of the internal magnetic field $\dot{\vect{p}} = \vect{v} \times \bar{b} \hat{z}$ with initial condition $\vect{p}(0) = \vect{q}$. Even in the presence of a magnetic field, the Fermi surface is a conserved quantity (i.e. $\dot{\epsilon}(\vect{q}) = 0$), and therefore only such states that live on the Fermi surface contribute to the DC conductivity of Eq.~\ref{semiclassicalcond}. The time-averaged group velocity for states on the Fermi surface is straightforward to compute. Let us parametrize a state on the Fermi surface as $\vect{p}(t) = k_F[\cos \phi(t) \hat{x} + \sin \phi(t) \hat{y}]$ with $\vect{p}(0) = \vect{q} = k_F[\cos \phi_0 \hat{x} + \sin \phi_0 \hat{y}]$. The group velocity is $\vect{v}(\vect{p}(t)) = v_F[\cos \phi(t) \hat{x} + \sin \phi(t) \hat{y}]$, and the solution to the equation of motion is $\phi(t) = \omega_c t+ \phi_0$ where we have now defined the cyclotron frequency $\omega_c = \frac{\bar{b} v_F}{k_F}$. The time-averaged group velocity is now easily seen to be
\beqn
\langle \vect{v}(\phi_0) \rangle &=& v_F\int_{-\infty}^0 \frac{dt}{\tau} \text{ } e^{t/\tau}[\cos \phi(t) \hat{x} + \sin \phi(t) \hat{y} ],\cr\cr
&=& \frac{v_F}{1 + \omega_c^2 \tau^2 }\bigg \{[ \omega_c \tau \sin \phi_0 + \cos \phi_0] \hat{x} \cr\cr &+&  [ \sin \phi_0 - \omega_c \tau \cos \phi_0  ] \hat{y}\bigg \}. \label{avevelresult}
\eeqn
The components of the DC conductivity can be read off from Eq.~\ref{semiclassicalcond} by plugging in Eq.~\ref{avevelresult} to take a standard Drude form. From this, the DC values of the CF conductivity tensor in valley-1 reads
\beqn
\sigma^\text{cf}_1 = \frac{\sigma^{\text{cf}}}{1 + \omega_c^2 \tau^2}
\begin{pmatrix}
    1 & \omega_c \tau\\
    - \omega_c \tau & 1
\end{pmatrix} 
\eeqn
with $\sigma^{\text{cf}} \sim v_F k_F \tau$. From this, the Hall resistivity of the CFs in each valley is $\rho^\text{cf}_{xy} = - \frac{\omega_c \tau}{\sigma^\text{cf}}$. After combining the two valleys together, and in the limit of $\sigma^s \ra \infty$, we can derive the DC values of the longitudinal and transverse components of the total electrical resistivity in Eq.~\ref{CantedCond} for the case of canted spin order. 

\section{Transformations of the gauge fields}

In this section we discuss the transformations of dynamical gauge fields $a$, $b$, as well as background gauge field $A$ under charge conjugation $\cC$ and time-reversal $\cT$. In Son's QED$_3$ CFL state, the gauge field $a$ is the dual of charge density, i.e. $\rho_e = \vect{\nabla} \times \vect{a}$ and $\psi$ is the vortex of the charge. Hence under time-reversal $\cT$, \beqn \cT a \cT^{-1} = a, \ \ \ \cT \psi \cT^{-1} = \psi^\dagger. \eeqn This is because the electron density does not change under $\cT$, but a charge vortex will become anti-vortex under $\cT$. Then under $\cC$, the charge density would change sign, and hence $\cC a \cC^{-1} = - a$. $\cC \cT$ together is an anti-unitary transformation which acts on $\psi$ in the same way as ordinary time-reversal operation on the electron operator~\cite{Son2015}. 

We have the freedom of choosing how $b$ transforms under $\cT$ and $\cC$. Our choice is \beqn \cT b \cT^{-1} = b, \ \ \ \cC b \cC^{-1} = b. \eeqn The background gauge field $A$ would transform as \beqn \cT A \cT^{-1} = - A, \ \ \ \cC A \cC^{-1} = - A. \eeqn Under these preparations, one can check that $\cL_1$ and $\cL_2$ in Eq.~\ref{LSon2} are connected to each other through $\cT$. 

Under $\cC \cT$, the three gauge fields transform as \beqn && (\cC \cT ) a (\cC \cT )^{-1} = -a, \ \ (\cC \cT ) b (\cC \cT )^{-1} = b, \cr \cr && (\cC \cT ) A (\cC \cT )^{-1} = A, \ \ (\cC \cT ) \psi (\cC \cT )^{-1} = \ii \sigma^y \psi. \eeqn 

let us perform the PH transformation on the Lagrangian of valley-2 in Eq.~\ref{LSon2}, which amounts to a $\mathcal{CT}$ transformation followed by inserting a background integer quantum hall state: \beqn \cL_2 &\ra& \cL'_{2} = \cL_{+1/2}(\psi_2, a_2) + \mu \psi^\dagger_2 \psi_2 - \frac{\ii}{2\pi} a_2 d b_2 \cr\cr &+& \frac{2\ii}{4\pi} b_2 d b_2 - \frac{\ii}{2\pi} b_2 d A_2 - \frac{\ii}{4\pi} c d c + \frac{\ii}{2\pi} c dA_2. \label{LSon3} \eeqn

One can check that $\cL_2$ in Eq.~\ref{LSon2} under $\cC \cT$ becomes $\cL'_2$ in Eq.~\ref{LSon3} without the last two terms. Now let us take $\cL_2'$ in Eq.~\ref{LSon3}, and perform a similar transformation~\cite{SEIBERG2016}, i.e. relabelling $b \ra b - c$ and $c \ra 2b - c$. $\cL_2'$ becomes \beqn \cL'_{2} &\ra& \cL_{+1/2}(\psi_2, a_2) + \mu \psi^\dagger_2 \psi_2 - \frac{\ii}{2\pi} a_2 d b_2 \cr\cr &-& \frac{2\ii}{4\pi} b_2 d b_2 + \frac{\ii}{2\pi} b_2 d A_2 + \frac{\ii}{4\pi} c d c + \frac{\ii}{2\pi} c da_2. \eeqn Integrating out gauge field $c$ would generate a CS term for $a_2$ with level $-1$: $- \frac{\ii}{4\pi} a_2 d a_2$, combining it with $\cL_{+ 1/2}(\psi_2, a_2)$ generates $\cL_{- 1/2}(\psi_2, a_2)$. Now $\cL'_2$ has returned to its original form $\cL_2$ in Eq.~\ref{LSon2}. Hence $\cL_2'$ describes the same state as $\cL_2$.

Hence the cCFL has different representations, $\cL = \cL_1 + \cL_2$, or $\cL = \cL_1 + \cL'_2$, where $\cL'_2$ is given by Eq.~\ref{LSon3}: \beqn \cL_{1} &=& \cL_{+1/2}(\psi_1, a_1) - \mu \psi^\dagger_1 \psi_1 + \frac{\ii}{2\pi} a_1 d b_1 \cr\cr &+& \frac{2\ii}{4\pi} b_1 d b_1 + \frac{\ii}{2\pi} b_1 d A_1 + \cdots; \cr\cr \cL'_2 &=& \cL_{+1/2}(\psi_2, a_2) + \mu \psi^\dagger_2 \psi_2 - \frac{\ii}{2\pi} a_2 d b_2 \cr\cr &+& \frac{2\ii}{4\pi} b_2 d b_2 - \frac{\ii}{2\pi} b_2 d A_2 - \frac{\ii}{4\pi} c d c + \frac{\ii}{2\pi} c dA_2 + \cdots. \label{LSon4} \eeqn  Now one can show that, in this new representation, when the CF exciton condensate of $\psi_1$ and $\psi_2$ forms a gapped band with trivial topology, the system would become a chiral spin liquid state described by Eq.~\ref{CSL}.

\end{document}